\documentclass[twocolumn, aps, pra, superscriptaddress, floatfix]{revtex4}
\usepackage[T1]{fontenc}
\usepackage[latin9]{inputenc}
\usepackage{color}
\usepackage{bm}
\usepackage{xcolor}
\usepackage{amsmath}
\usepackage{amssymb}
\usepackage{graphicx}
\usepackage[sort&compress]{natbib}
\usepackage{braket}
\usepackage{psfrag}
\usepackage{tikz}
\usepackage[normalem]{ulem}
\usepackage{qcircuit}
\usepackage{subcaption}
\usepackage{accents}
\usepackage{cleveref}
\usepackage{soul}
\usepackage{multirow}

\newcommand{\R}{\mathbb{R}} 

\DeclareMathOperator{\EX}{\mathbb{E}}

\begin{document}

\title{Quantum Generative Adversarial Networks \\
for Learning and Loading Random Distributions}

\author{Christa Zoufal}%
\email{ouf@zurich.ibm.com}
\affiliation{IBM Research -- Zurich, Rueschlikon 8803, Switzerland}
\affiliation{ETH Zurich, Zurich 8092, Switzerland}

\author{Aur\'{e}lien Lucchi}
\affiliation{ETH Zurich, Zurich 8092, Switzerland}

\author{Stefan Woerner}
\affiliation{IBM Research -- Zurich, Rueschlikon 8803, Switzerland}
		
\date{\today}

\begin{abstract}
Quantum algorithms have the potential to outperform their classical counterparts in a variety of tasks. The realization of the advantage often requires the ability to load classical data efficiently into quantum states. However, the best known methods require $\mathcal{O}\left(2^n\right)$ gates to load an exact representation of a generic data structure into an $n$-qubit state. This scaling can easily predominate the complexity of a quantum algorithm and, thereby, impair potential quantum advantage. 

Our work presents a hybrid quantum-classical algorithm for efficient, approximate quantum state loading. 
More precisely, we use quantum Generative Adversarial Networks (qGANs) to facilitate efficient learning and loading of generic probability distributions -- implicitly given by data samples -- into quantum states.
Through the interplay of a quantum channel, such as a variational quantum circuit, and a classical neural network, the qGAN can learn a representation of the probability distribution underlying the data samples and load it into a quantum state. 

The loading requires $\mathcal{O}\left(poly\left(n\right)\right)$ gates and can, thus, enable the use of potentially advantageous quantum algorithms, such as Quantum Amplitude Estimation.

We implement the qGAN distribution learning and loading method with Qiskit and test it using a quantum simulation as well as actual quantum processors provided by the IBM Q Experience.
Furthermore, we employ quantum simulation to demonstrate the use of the trained quantum channel in a quantum finance application.

\end{abstract}

\maketitle


\section{Introduction}

The realization of many promising quantum algorithms is impeded by the assumption that data can be efficiently loaded into a quantum state \cite{hhl, qPCA, nielsen, brassard}. However, this may only be achieved for particular but not for generic data structures. In fact, data loading can easily dominate the overall complexity of an otherwise advantageous quantum algorithm \cite{Aaronson2015_finePrint}.
In general, data loading relies on the availability of a quantum state preparing channel.
But, the exact preparation of a generic state in $n$ qubits requires $\mathcal{O}\left(2^n\right)$ gates \cite{Grover2000SynthesisSuperpos, Sanders2019StatePrep, Plesch2010StatePrep, Shende:2005:SQL:1120725.1120847}.
In many cases, this complexity diminishes a potential quantum advantage.

This work discusses the training of an approximate, efficient data loading channel with Quantum Machine Learning for particular data structures. More specifically, we present a feasible learning and loading scheme for generic probability distributions based on a generative model. The scheme utilizes a hybrid quantum-classical implementation of a Generative Adversarial Network (GAN) \cite{goodfellow, Kurach2018TheGL} to train a quantum channel such that it reflects a probability distribution implicitly given by data samples.

In classical machine learning, GANs have proven useful for generative modeling. These algorithms employ two competing neural networks - a generator and a discriminator - which are trained alternately. Replacing either the generator, the discriminator, or both with quantum systems translates the framework to the quantum computing context \cite{lloyd2}. 

The first theoretic discussion of quantum GANs (qGANs) was followed by demonstrations of qGAN implementations.
Some focus on quantum state estimation \cite{Paris2010_QuantumStateEstimation}, i.e.~finding a quantum channel whose output is an estimate to a given quantum state \cite{killoran, benedetti, hu}. 
Others exploit qGANs to generate classical data samples in accordance with the training data's underlying distribution \cite{zheng,romero,LearnandInference}.

In contrast, our qGAN implementation learns and loads probability distributions into quantum states.
More specificially, the aim of the qGAN is not to produce classical samples in accordance with given classical training data but to train the quantum generator to create a quantum state which represents the data's underlying probability distribution.
The resulting quantum channel, given by the quantum generator, enables efficient loading of an approximated probability distribution into a quantum state. It can be easily prepared and reused as often as needed. 
Now, applying this qGAN scheme for data loading can facilitate quantum advantage in combination with other algorithms such as Quantum Amplitude Estimation (QAE) \cite{brassard} or the HHL-algorithm \cite{hhl}. Notably, QAE and HHL -- given a well-conditioned matrix and a suitable classical right-hand-side \cite{Aaronson2015_finePrint} -- are both compatible with approximate state preparation as these algorithms are stable to small errors in the input state, i.e.~small deviations in the input only lead to small deviations in the result.

The remainder of this paper is structured as follows.
Sec.~\ref{sec:gan} explains classical GANs.
Then, the qGAN-based distribution learning and loading scheme is introduced and analyzed on different test cases in Sec.~\ref{sec:qGANs}.
In Sec.~\ref{sec:results}, we discuss the exploitation of qGANs to facilitate quantum advantage in financial derivative pricing: First, we discuss the training of the qGAN with data samples drawn from a log-normal distribution and present the results obtained with a quantum simulator and the IBM Q Boeblingen superconducting quantum computer with 20 qubits, both accessible via the IBM Q Experience \cite{ibmQX}. Then, the resulting quantum channel is used in combination with QAE to price a European call option.
Finally, Sec.~\ref{sec:discussion} presents the conclusions and a discussion on open questions and additional possible applications of the presented scheme.

\section{Generative Adversarial Networks} \label{sec:gan}

The generative models considered in this work, GANs \cite{goodfellow, Kurach2018TheGL}, employ two neural networks - a generator and a discriminator - to learn random distributions that are implicitly given by training data samples.

Originally, GANs have been used in the context of image generation and modification. In contrast to previously used generative models, such as Variational Auto Encoders (VAEs) \cite{Kingma2013AutoEncodingVB, Burda2015ImportanceWA}, GANs managed to generate sharp images and consequently gained popularity in the machine learning community \cite{Dumoulin2017AdversLearnedInference}. 
VAEs and other generative models relying on log-likelihood optimization are prone to generating blurry images. 
Particularly for multi-modal data, log-likelihood optimization tends to spread the mass of a learned distribution over all modes. GANs, on the other hand, tend to focus the mass on each mode \cite{goodfellow, Metz2016UnrolledGA}.

\begin{figure}[h!]
\captionsetup{singlelinecheck = false, format= hang, justification=raggedright, font=footnotesize, labelsep=space}
\begin{center}
\includegraphics[width=\linewidth]{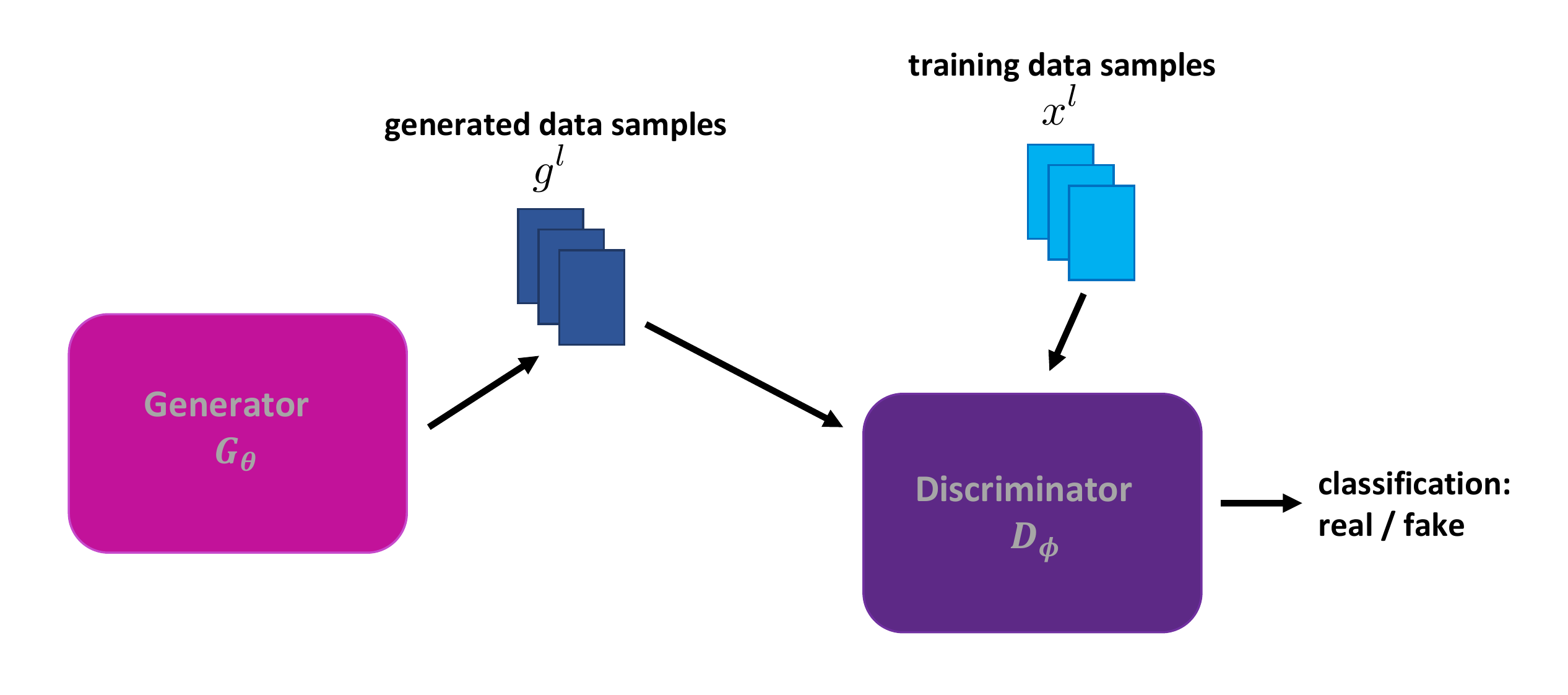}
\end{center}
\caption{Generative Adversarial Network: First, the generator creates data samples which shall be indistinguishable from the training data. Second, the discriminator tries to differentiate between the generated samples and the training samples. The generator and discriminator are trained alternately.}
\label{fig:gan}
\end{figure}

Suppose a classical training data set $X=\set{x^0, \ldots, x^{s-1}} \subset \mathbb{R}^{k_{out}}$ sampled from an unknown probability distribution $p_{\text{real}}$. Let $G_{\theta}: \mathbb{R}^{k_{in}} \rightarrow \mathbb{R}^{k_{out}}$ and $D_{\phi}:\mathbb{R}^{k_{out}} \rightarrow \{0, 1\}$ denote the generator and the discriminator networks, respectively. The corresponding network parameters are given by $\theta \in \R^{k_g}$ and $\phi \in \R^{k_d}$. 
The generator $G_{\theta}$ translates samples from a fixed prior distribution $p_{\text{prior}}$ in $\mathbb{R}^{k_{in}}$ into samples which shall be indistinguishable from samples of the real distribution $p_{\text{real}}$ in $\mathbb{R}^{k_{out}}$. 
The discriminator $D_{\phi}$, on the other hand, tries to distinguish between data from the generator and from training set.
The training process is illustrated in Fig.~\ref{fig:gan}. 
	
The optimization objective of classical GANs may be defined in various ways. In this work, we consider the non-saturating loss \cite{fedus2018many} which is also used in the code of the original GAN paper \cite{goodfellow}. The generator's loss function

\begin{equation}
\begin{split}
	L_G\left(\phi,\theta\right) = \thinspace -\mathbb{E}_{z\sim p_{\text{prior}}}\left[\log\left(D_{\phi}\left(G_{\theta}\left(z\right)\right)\right)\right]
	\end{split}
\end{equation}

aims at maximizing the likelihood that the generator creates samples that are labeled as real data samples.
On the other hand, the discriminator's loss function
\begin{equation}
\begin{split}
	&L_D\left(\phi,\theta\right) = \\
 &\mathbb{E}_{x\sim p_{\text{real}}}\left[\log D_{\phi}\left(x\right) \right] +
	\mathbb{E}_{z\sim p_{\text{prior}}}\left[\log\left(1-D_{\phi}\left(G_{\theta}\left(z\right)\right)\right)\right]
	\end{split}
\end{equation}

aims at maximizing the likelihood that the discriminator labels training data samples as training data samples and generated data samples as generated data samples.

In practice, the expected values are approximated by batches of size $m$
\begin{equation}
\begin{split}
	L_G\left(\phi,\theta\right) = -\frac{1}{m}\sum\limits_{l=1}^{m}\left[\log\left(D_{\phi}\left(G_{\theta}\left(z^{l}\right)\right)\right)\right] \text{, and}
	\end{split}
\end{equation}
\begin{equation}
\begin{split}
	&L_D\left(D_{\phi}, G_{\theta}\right) = \\
	&\frac{1}{m}\sum\limits_{l=1}^{m}\left[\log D_{\phi}\left(x^{l}\right)  + \right.
	\left.\log\left(1-D_{\phi}\left(G_{\theta}\left(z^{l}\right)\right)\right)\right],
	\end{split}
\end{equation}
for $x^l \in X$ and $z^l \sim p_{\text{prior}}$.

Training the GAN is equivalent to searching for a Nash-equilibrium of a two-player game:
\begin{align}
 \label{eq:minmaxGenerator}
\underset{\theta}{\max}&\: L_G\left(\phi,\theta\right) \\
\label{eq:minmax}
 \underset{\phi}{\max}&\: L_D\left(\phi,\theta\right).
\end{align}

Typically, the optimization of Eq.~\eqref{eq:minmaxGenerator} and Eq.~\eqref{eq:minmax} employs alternating update steps for the generator and the discriminator. These alternating steps lead to non-stationary objective functions, i.e.~an update of the generator's (discriminator's) network parameters also changes the discriminator's (generator's) loss function.
Common choices to perform the update steps are ADAM \cite{adam} and AMSGRAD \cite{amsgrad}, which are adaptive-learning-rate, gradient-based optimizers that use an exponentially decaying average of previous gradients, and are well suited for solving non-stationary objective functions \cite{adam}.

\section{qGAN Distribution Learning} \label{sec:qGANs}

Our qGAN implementation uses a \textbf{quantum} generator and a \textbf{classical} discriminator to capture the probability distribution of \textbf{classical} training samples.

Notably, the aim of this approach is to train a data loading quantum channel for generic probability distributions. As discussed in Sec.~\ref{sec:gan}, GAN-based learning is explicitly suitable to capture not only uni-modal but also multi-modal distributions, as we will also demonstrate later in this section.

In this setting, a parametrized quantum channel, i.e.~the quantum generator, is trained to transform a given $n$-qubit input state $\ket{\psi_{\text{in}}}$ to an $n$-qubit output state  
\begin{equation}
G_{\theta}\ket{\psi_{\text{in}}} = \ket{g_{\theta}} = \sum\limits_{j=0}^{2^n-1}\sqrt{p_{\theta}^{j}}\ket{j},
\end{equation}
where $p_{\theta}^{j}$ describe the resulting occurrence probabilities of the basis states $\ket{j}$.

For simplicity, we now assume that the domain of $X$ is $\{0, ..., 2^n-1\}$ and, thus, the existence of a natural mapping between the sample space of the training data and the states that can be represented by the generator.
This assumption can be easily relaxed, for instance, by introducing an affine mapping between $\{0, ..., 2^n-1\}$ and an equidistant grid suitable for $X$.
In this case, it might be necessary to map points in $X$ to the closest grid point to allow for an efficient training.
The number of qubits $n$ determines the distribution loading scheme's resolution, i.e.~the number of discrete values $2^n$ that can be represented.
During the training, this affine mapping can be applied classically after measuring the quantum state. However, when the resulting quantum channel is used within another quantum algorithm the mapping must be executed as part of the quantum circuit.
As was discussed in \cite{wor}, such an affine mapping can be implemented in a gate-based quantum circuit with linearly many gates. 

\begin{figure}[h!]
\captionsetup{singlelinecheck = false, format= hang, justification=raggedright, font=footnotesize, labelsep=space} 
   \centering{
    \includegraphics[width=\linewidth]{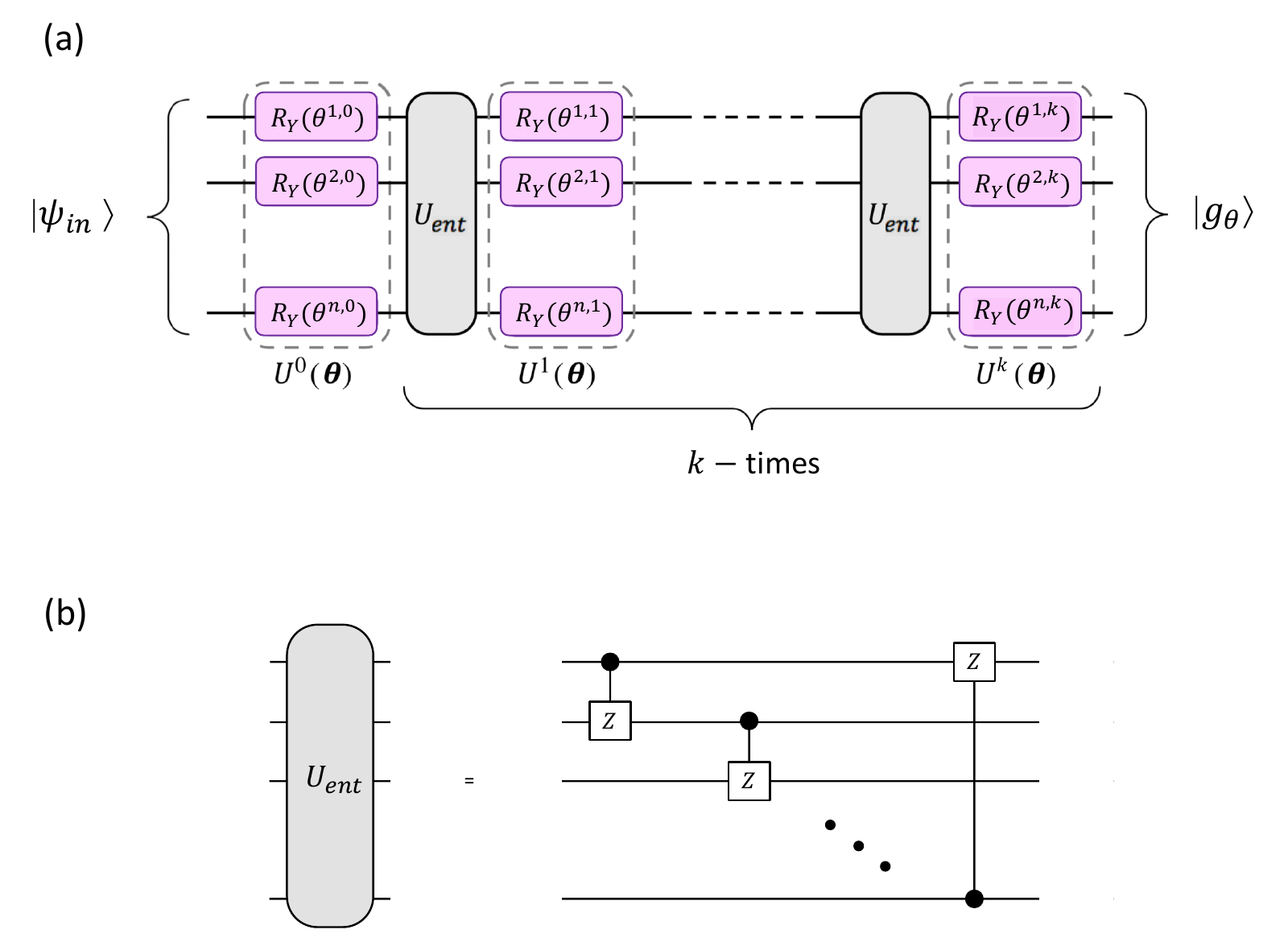}}
\caption{The variational form, depicted in (a), with depth $k$ acts on $n$ qubits. It is composed of $k+1$ layers of single-qubit Pauli-$Y$-rotations and $k$ entangling blocks $U_{\text{ent}}$. As illustrated in (b), each entangling block applies $CZ$ gates from qubit $i$ to qubit $\left(i+1\right) \mod\: n,\:i \in \set{0, \ldots, n-1}$ to create entanglement between the different qubits.}
 \label{fig:varForm}
\end{figure}

The quantum generator is implemented by a variational form \cite{Aspuru2015_VariationalCircuits}, i.e.~a parametrized quantum circuit.
We consider variational forms consisting of alternating layers of parametrized single-qubit rotations, here Pauli-Y-rotations $\left(R_Y\right)$ \cite{nielsen}, and blocks of two-qubit gates, here controlled-$Z$-gates $\left(CZ\right)$ \cite{nielsen}, called entanglement blocks $U_{\text{ent}}$.
The circuit consists of a first layer of $R_Y$ gates, and then $k$ alternating repetitions of $U_{\text{ent}}$ and further layers of $R_Y$ gates.
The rotation acting on the $i^{\text{th}}$ qubit in the $j^{\text{th}}$ layer is parametrized by $\theta^{i,j}$.
Moreover, the parameter $k$ is called the depth of the variational circuit.
If such a variational circuit acts on $n$ qubits it uses in total $(k+1)n$ parametrized single-qubit gates and $kn$ two-qubit gates, see Fig.~\ref{fig:varForm} for an illustration.
Similarly to increasing the number of layers in deep neural networks \cite{Goodfellow2016_DeepL}, increasing the depth $k$ enables the circuit to represent more complex structures and increases the number of parameters. 
Another possibility to increase the quantum generator's ability to represent complex correlations is adding ancilla qubits as this facilitates an isometric instead of a unitary mapping \cite{nielsen}, see Appendix~\ref{app:isometry} for more details.

The rationale behind choosing a variational form with $R_Y$ and $CZ$ gates, e.g.~in contrast to other Pauli rotations and two-qubit gates, is that for $\theta^{i,j} = 0$ the variational form does not have any effect on the state amplitudes but only flips the phases. 
These phase flips do not perturb the modeled probability distribution which solely depends on the state amplitudes.
Thus, if a suitable $\ket{\psi_{\text{in}}}$ can be loaded efficiently, the variational form allows its exploitation.

To train the qGAN, samples are drawn by measuring the output state $\ket{g_{\theta}}$ in the computational basis, where the set of possible measurement outcomes is $\ket{j},\:j \in \set{0, \ldots, 2^n-1}$. 
Unlike in the classical case, the sampling does not require a stochastic input but is based on the inherent stochasticity of quantum measurements. 
Notably, the measurements return classical information, i.e.~$p_j$ being defined as the measurement frequency of $\ket{j}$.

The scheme can be easily extended to $d$-dimensional distributions by choosing $d$ qubit registers with $n_i$ qubits each, for $i = 1, \ldots, d$, and constructing a multi-dimensional grid, see Appendix~\ref{app:multi} for an explicit example of a qGAN trained on multivariate data.

A carefully chosen input state $\ket{\psi_{\text{in}}}$ can help to reduce the complexity of the quantum generator and the number of training epochs as well as to avoid local optima in the quantum circuit training.
Since the preparation of $\ket{\psi_{\text{in}}}$ should not dominate the overall gate complexity, the input state must be loadable with $\mathcal{O}\left(poly\left(n\right)\right)$ gates.
This is feasible, e.g.~for efficiently integrable probability distributions, such as log-concave distributions \cite{grover}.
In practice, statistical analysis of the training data can guide the choice for a suitable $\ket{\psi_{\text{in}}}$ from the family of efficiently loadable distributions, e.g.~by matching expected value and variance.

In Sec.~\ref{subsec:benchmarking}, we present a broad simulation study that analyzes the impact of $\ket{\psi_{\text{in}}}$ as well as the circuit depth $k$.

The classical discriminator, a standard neural network consisting of several layers that apply non-linear activation functions, processes the data samples and labels them either as being real  or generated.

Notably, the topology of the networks, i.e.~number of nodes and layers, needs to be carefully chosen to ensure that the discriminator does not overpower the generator and vice versa.

Given $m$ data samples $g^l$ from the quantum generator and $m$ randomly chosen training data samples $x^{l}$, where $l = 1, \ldots, m$, the loss functions of the qGAN are
\begin{align}
\begin{split}
\label{eq:lossqGanG}
	 L_G\left(\phi, \theta\right) = -\frac{1}{m}\sum\limits_{l=1}^{m}\left[\log D_{\phi}\left(g^{l}\right) \right],
	\end{split}
\end{align}
for the generator, and
\begin{align}
\begin{split}
\label{eq:lossqGanD}
	& L_D\left(\phi, \theta\right) = \\
	&\frac{1}{m}\sum\limits_{l=1}^{m}\left[\log D_{\phi}\left(x^{l}\right)  + \right. 
	\left. \log\left(1-D_{\phi}\left(g^{l}\right)\right)\right],
	\end{split}
\end{align}
for the discriminator, respectively.

As in the classical case, see Eq.~(\ref{eq:minmaxGenerator}) and (\ref{eq:minmax}), the loss functions are optimized alternately with respect to the generator's parameters $\theta$ and the discriminator's parameters $\phi$.

\subsection{Simulation Study}  \label{subsec:benchmarking}
Next, we present the results of a broad simulation study on training qGANs with different settings for different target distributions.

The quantum generator is implemented with Qiskit \cite{qiskit} which enables the circuit execution with quantum simulators as well as quantum hardware provided by the IBM Q Experience \cite{ibmQX}.

We consider a quantum generator acting on $n=3$ qubits, which can represent $2^3=8$ values, namely $\{0, 1, \ldots, 7\}$.
We applied the method for $20,000$ samples of, first, a log-normal distribution with $\mu = 1$ and $\sigma = 1$, second, a triangular distribution with lower limit $l=0$, upper limit $u=7$ and mode $m=2$, and last, a bimodal distribution consisting of two superimposed Gaussian distributions with $\mu_1=0.5$, $\sigma_1=1$ and $\mu_2=3.5$, $\sigma_2=0.5$, respectively.
All distributions were truncated to $\left[0, 7\right]$ and the samples were rounded to integer values.

The generator's input state $\ket{\psi_{\text{in}}}$ is prepared according to a discrete uniform distribution, a truncated and discretized normal distribution with $\mu$ and $\sigma$ being empirical estimates of mean and standard deviation of the training data samples, or a randomly chosen initial distribution.

Preparing a uniform distribution on $3$ qubits requires the application of $3$ Hadamard gates, i.e.~one per qubit \cite{nielsen}. Loading a normal distribution involves more advanced techniques, see Appendix~\ref{app:normalInit} for further details.
For both cases, we sample the generator parameters from a uniform distribution on $[-\delta, +\delta]$, for $\delta = 10^{-1}$.
By construction of the variational form, the resulting distribution will be close to $\ket{\psi_{\text{in}}}$ but slightly perturbed. Adding small random perturbations helps to break symmetries and can, thus, help to  improve the training performance \cite{LEHTOKANGAS1998265, Thimm_percept97, Chen2019}.
To create a randomly chosen distribution, we set $\ket{\psi_{\text{in}}}=\ket{0}^{\otimes 3}$ and initialize the parameters of the variational form following a uniform distribution on $[-\pi, \pi]$.

From now on, we refer to these three cases as \emph{uniform}, \emph{normal}, and \emph{random} initialization.
.
Furthermore, we tested quantum generators with depths $k \in \set{1, 2, 3}$.

The discriminator, a classical neural network, is implemented with PyTorch \cite{pytorch}.
The neural network consists of a $50$-node input layer, a $20$-node hidden-layer and a single-node output layer. 
First, the input and the hidden layer apply linear transformations followed by Leaky ReLU functions \cite{goodfellow, Pedamonti_2018_Non-linear_Activation_Functions_NN, HeRectifiers15}.
Then, the output layer implements another linear transformation and applies a sigmoid function.
The network should neither be too weak nor too powerful to ensure that neither the generator nor the discriminator overpowers the other network during the training.
The used discriminator topology has been chosen based on empirical tests.

The qGAN is trained using AMSGRAD \cite{amsgrad} with the initial learning rate being $10^{-4}$.
Due to the utilization of first and second momentum terms, this is a robust optimization technique for non-stationary objective functions as well as for noisy gradients \cite{adam}, which makes it particularly suitable for running the algorithm on real quantum hardware.
Methods for the analytic computation of the quantum generator loss function's gradients are discussed in Appendix~\ref{app:gradients}.
The training stability is improved further by applying a gradient penalty on the discriminator's loss function \cite{Kodali2017OnCA, Roth2017StabilizingTO}.

In each training epoch, the training data is shuffled and split into batches of size $2,000$.
The generated data samples are created by preparing and measuring the quantum generator $2,000$ times.
Then, the batches are used to update the parameters of the discriminator and the generator in an alternating fashion. After the updates are completed for all batches, a new epoch starts.

According to the classical GAN literature, the loss functions do not neccessarily reflect whether the method converges \cite{Grnarova2018EvaluatingGV}.
In the context of training a quantum representation of some training data's underlying random distribution, the Kolmogorov-Smirnov statistic as well as the relative entropy represent suitable measures to evaluate the training performance.
Given the null-hypothesis that the probability distribution from $\ket{g_{\theta}}$ is equivalent to the probability distribution underlying $X$, the Kolmogorov-Smirnov statistic $D_{KS}$ determines whether the null-hypothesis is accepted or rejected with a certain confidence level, here set to $95\%$.
The relative entropy quantifies the difference between two probability distributions.
In the following, we analyze the results using these two statistical measures, which are formally introduced in Appendix~\ref{app:statMeas}.

\begin{figure}[h!]
\captionsetup{singlelinecheck = false, format= hang, justification=raggedright, font=footnotesize, labelsep=space}
\begin{center}
\includegraphics[width=\linewidth]{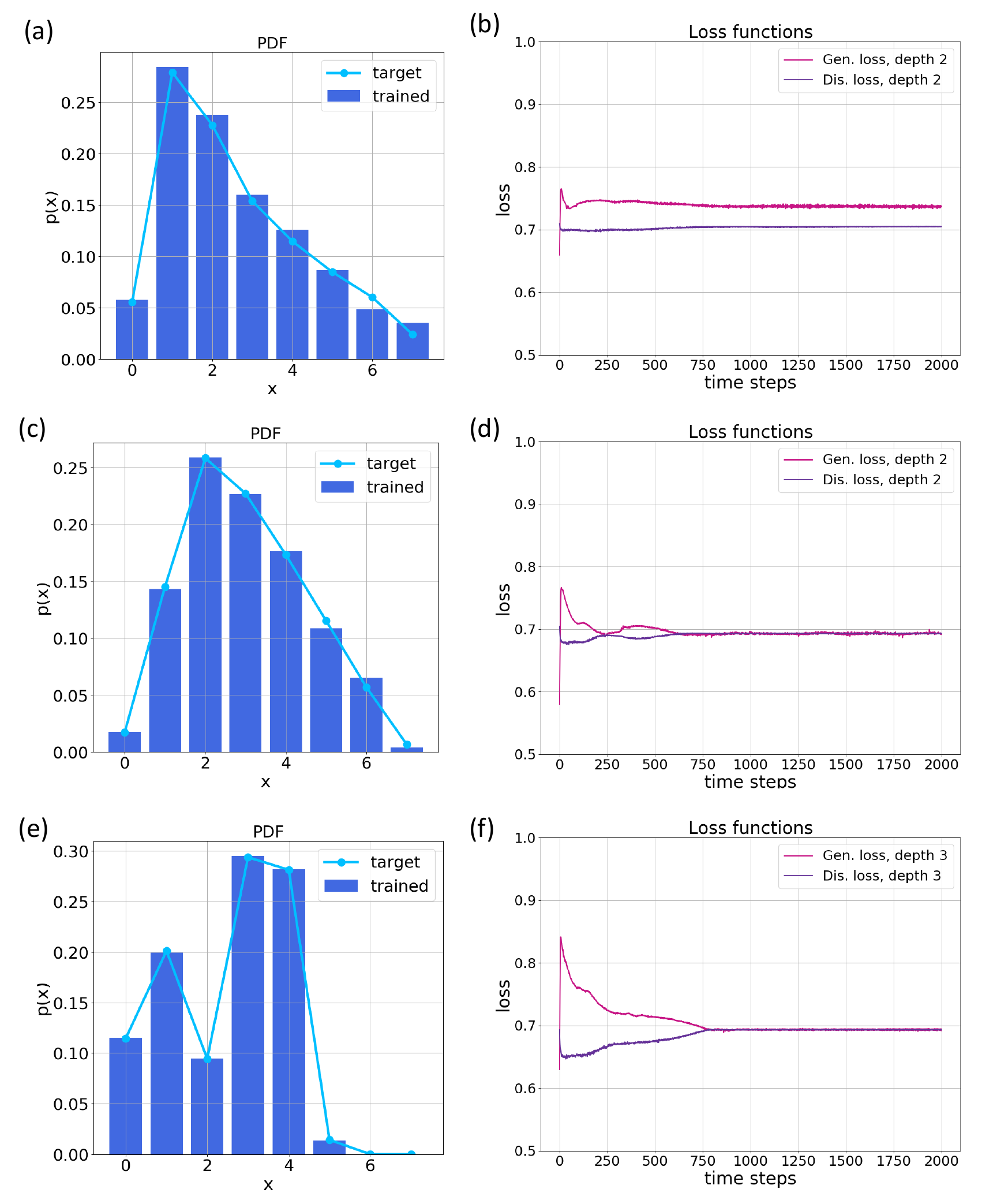}
\end{center}
\caption{Result of training the qGAN for a log-normal target distribution with normal initialization and a depth $2$ generator (a, b), a triangular target distribution with random initialization and a depth $2$ generator (c, d), and a bimodal target distribution with uniform initialization and a depth $3$ generator (e, f).
The presented probability density functions correspond to the trained $\ket{g_{\theta}}$  (a, c, e) and the loss function progress is illustrated for the generator as well as for the discriminator (b, d, f).
}
\label{fig:benchmark}
\end{figure}

For each setting, we repeat the training $10$ times to get a better understanding of the robustness of the results.
Table~\ref{tbl:trainingBenchmarking} shows aggregated results over all 10 runs and presents the mean $\mu_{KS}$, the standard deviation $\sigma_{KS}$ and the number of accepted runs $n_{\leq b}$  according to the Kolmogorov-Smirnov statistic as well as the mean $\mu_{RE}$ and standard deviation $\sigma_{RE}$ of the relative entropy outcomes between the generator output and the corresponding target distribution.
The data shows that increasing the quantum generator depth $k$ usually improves the training outcomes.
Furthermore, the table illustrates that a carefully chosen initialization can have favorable effects, as can be seen especially well for the bimodal target distribution with normal initialization.
Since the standard deviations are relatively small and the number of accepted results is usually close to 10, at least for depth $k \geq 2$, we conclude that the presented approach is quite robust and applicable also to more complicated distributions.
Fig.~\ref{fig:benchmark} illustrates the results for one example of each target distributions.
 
\begin{table}[h]
\captionsetup{singlelinecheck = false, format= hang, justification=raggedright, font=footnotesize, labelsep=space}
\begin{tabular}{|c|c|c|c|c|c|c|c|}
\hline
data& 	init & $k$	& $\mu_{KS}$ &	$\sigma_{KS}$	& $n_{\leq b}$	& $\mu_{RE}$	& $\sigma_{RE}$ \\
\hline
\hline
\multirow{9}{* } {log-normal}& \multirow{3}{*}{uniform} & 1	& 0.0522	& 0.0214 &	9	&0.0454	&0.0856\\
 & & 	2 &	0.0699  & 0.0204	 & 7	 & 0.0739	 & 0.0510 \\
  & &	3 &0.0576	 &0.0206	 &9	 &0.0309	 &0.0206 \\

 & \multirow{3}{*}{normal}  &	1	 &0.1301	 &0.1016 &	5	 &0.1379 &	0.1449\\
 & &2	 &0.1380 &	0.0347	 &1 &	0.1283	 &0.0716\\
 & &	3	 &0.0810 &	0.0491	 & 7 &	0.0435 &	0.0560 \\
 &\multirow{3}{*}{random}	 &1	 &0.0821	 &0.0466	& 7	 &0.0916 &	0.0678\\
 & &	2	 &0.0780 &	0.0337	 &6	 &0.0639	 &0.0463\\
 & &	3	 &0.0541 &	0.0174 &	10 &	0.0436	 &0.0456\\
 \hline
\multirow{9}{* } {triangular	}&\multirow{3}{*}{uniform}&	1&	0.0880&	0.0632	&6	&0.0624&	0.0535\\
&&	2&	0.0336	&0.0174	&10&	0.0091&	0.0042\\
&&	3	&0.0695	&0.1028&	9	&0.0760 &	0.1929\\
&\multirow{3}{*}{normal} 	&1&	0.0288	&0.0106	&10	&0.0038&	0.0048\\
&&	2&	0.0484	&0.0424&	9	&0.0210&0.0315\\
&&	3&	0.0251	&0.0067& 10&	0.0033	&0.0038 \\
&\multirow{3}{*}{random}	 &	1	&0.0843&	0.0635	&7&0.1050	&0.1387\\
&&	2&0.0538&	0.0294	&9&	0.0387	&0.0486\\
&&	3	&0.0438	&0.0163&	10	&0.0201	&0.0194\\
\hline
\multirow{9}{* } {bimodal}&\multirow{3}{*}{uniform}	&1	&0.1288	&0.0259	&0&0.3254	&0.0146\\
&&	2	&0.0358	&0.0206	&10	&0.0192&	0.0252\\
&&	3	&0.0278	&0.0172&	10	&0.0127	&0.0040\\
&\multirow{3}{*}{normal} 	&1&	0.0509	&0.0162	&9&	0.3417&	0.0031\\
&&	2	&0.0406	&0.0135	&10	&0.0114&	0.0094\\
&&	3	&0.0374&	0.0067	&10	&0.0018&0.0041\\
&\multirow{3}{*}{random}		&1	&0.2432&	0.0537	&0	&0.5813&	0.2541\\
&&	2	&0.0279&0.0078	&10	&0.0088&	0.0060\\
&&	3 &	0.0318&0.0133&	10&	0.0070&	0.0069\\
\hline
\end{tabular}
\caption{Benchmarking the qGAN training for log-normal, triangular and bimodal target distributions, uniform, normal and random initializations, and variational circuits with depth $1, 2$ and $3$.
The tests were repeated $10$ times using quantum simulation. 
The table shows the mean $\left(\mu\right)$ and the standard deviation $\left(\sigma\right)$ of the Kolmogorov-Smirnov statistic $\left(KS\right)$ as well as of the relative entropy $\left(RE\right)$ between the generator output and the corresponding target distribution.
Furthermore, the table shows the number of runs accepted according to the Kolmogorov Smirnov statistic $\left(n_{\leq b}\right)$ with confidence level $95\%$, i.e., with acceptance bound $b=0.0859$.}
\label{tbl:trainingBenchmarking}
\end{table}
 
\section{Application in Quantum Finance} \label{sec:results}

Now, we demonstrate that training a data loading unitary with qGANs can facilitate financial derivative pricing.
More precisely, we employ qGANs to learn and load a model for the spot price of an asset underlying a European call option. We perform the training for different initial states with a quantum simulator, and also execute the learning and loading method for a random initialization on an actual quantum computer, the IBM Q Boeblingen 20 qubit chip. 
Then, the fair price of the option is estimated by sampling from the resulting distribution, as well as with a QAE algorithm \cite{brassard, wor} that uses the quantum generator trained with IBM Q Boeblingen for data loading.
A detailed description of the QAE algorithm is given in Appendix~\ref{app:QAE}.

The owner of a European call option is permitted, but not obliged, to buy an underlying asset for a given strike price $K$ at a predefined future maturity date $T$, where the asset's spot price at maturity $S_T$ is assumed to be uncertain.
If $S_T \leq K$, i.e.~the spot price is below the strike price, it is unreasonable to exercise the option and there is no payoff.
However, if $S_T > K$, exercising the option to buy the asset for price $K$ and immediately selling it again for $S_T$ can realize a payoff $S_T - K$. Thus, the payoff of the option is defined as $\max\lbrace S_T - K, 0 \rbrace$.
Now, the goal is to evaluate the expected payoff $\EX\left[\max\lbrace S_T - K,0 \rbrace\right]$, whereby $S_T$ is assumed to follow a particular random distribution.
This corresponds to the fair option price before discounting \cite{BlackScholes}. Here, the discounting is neglected to simplify the problem.

To demonstrate and verify the applicability of the suggested training method, we implement a small illustrative example that is based on the analytically computable standard model for European option pricing, the Black-Scholes model \cite{BlackScholes}. The qGAN algorithm is used to train a corresponding data loading unitary which enables the evaluation of characteristics of this model, such as the expected payoff, with QAE.

It should be noted that the Black-Scholes model often over-simplifies the real circumstances.
In more realistic and complex cases, where the spot price follows a more generic stochastic process or where the payoff function has a more complicated structure, options are usually evaluated with Monte Carlo simulations \cite{Glasserman2003}.
A Monte Carlo simulation uses $N$ random samples drawn from the respective distribution to evaluate an estimate for a characteristic of the distribution, e.g.~the expected payoff. The estimation error of this technique behaves like $\epsilon=\mathcal{O}(1/\sqrt{N})$.
When using $n$ evaluation qubits to run a QAE, this induces the evaluation of $N=2^n$ quantum samples to estimate the respective distribution charactersitic. Now, this quantum algorithm achieves a Grover-type error scaling for option pricing, i.e.~$\epsilon=\mathcal{O}(1/N)$ \cite{brassard, wor, Rebentrost2018}.
To evaluate an option's expected payoff with QAE, the problem must be encoded into a quantum operator that loads the respective probability distribution and implements the payoff function.

In this work, we demonstrate that this distribution can be loaded approximately by training a qGAN algorithm. 

In the remainder of this section, we first illustrate the training of a qGAN using classical quantum simulation.
Then, the results from running a qGAN training on actual quantum hardware are presented
Finally, we employ the generator trained with a real quantum computer to conduct QAE-based option pricing.

\subsection{QGAN Training} \label{subsec:lognormal}

According to the Black-Scholes model \cite{BlackScholes}, the spot price at maturity $S_T$ for a European call option is log-normally distributed.
Thus, we assume that $p_{\text{real}}$, which is typically unknown, is given by a log-normal distribution and generate the training data $X$ by randomly sampling from a log-normal distribution.

As in Sec.~\ref{subsec:benchmarking}, the training data set $X$ is constructed by drawing $20,000$ samples from a log-normal distribution with mean $\mu=1$ and standard deviation $\sigma=1$ truncated to $\left[0, 7\right]$, and then, rounding the sampled values to integers, i.e.~to the grid that can be natively represented by the generator.
We discuss a detailed analysis of training a model for this distribution with a depth $k=1$ quantum generator, which is sufficient for this small example, with different initializations, namely uniform, normal and random.
The discriminator and generator network architectures as well as the optimization method are chosen equivalently to the ones described in Sec.~\ref{subsec:benchmarking}.

\begin{figure}[h!]
\captionsetup{singlelinecheck = false, format= hang, justification=raggedright, font=footnotesize, labelsep=space} 
\centering{
	\includegraphics[width=0.45\textwidth]{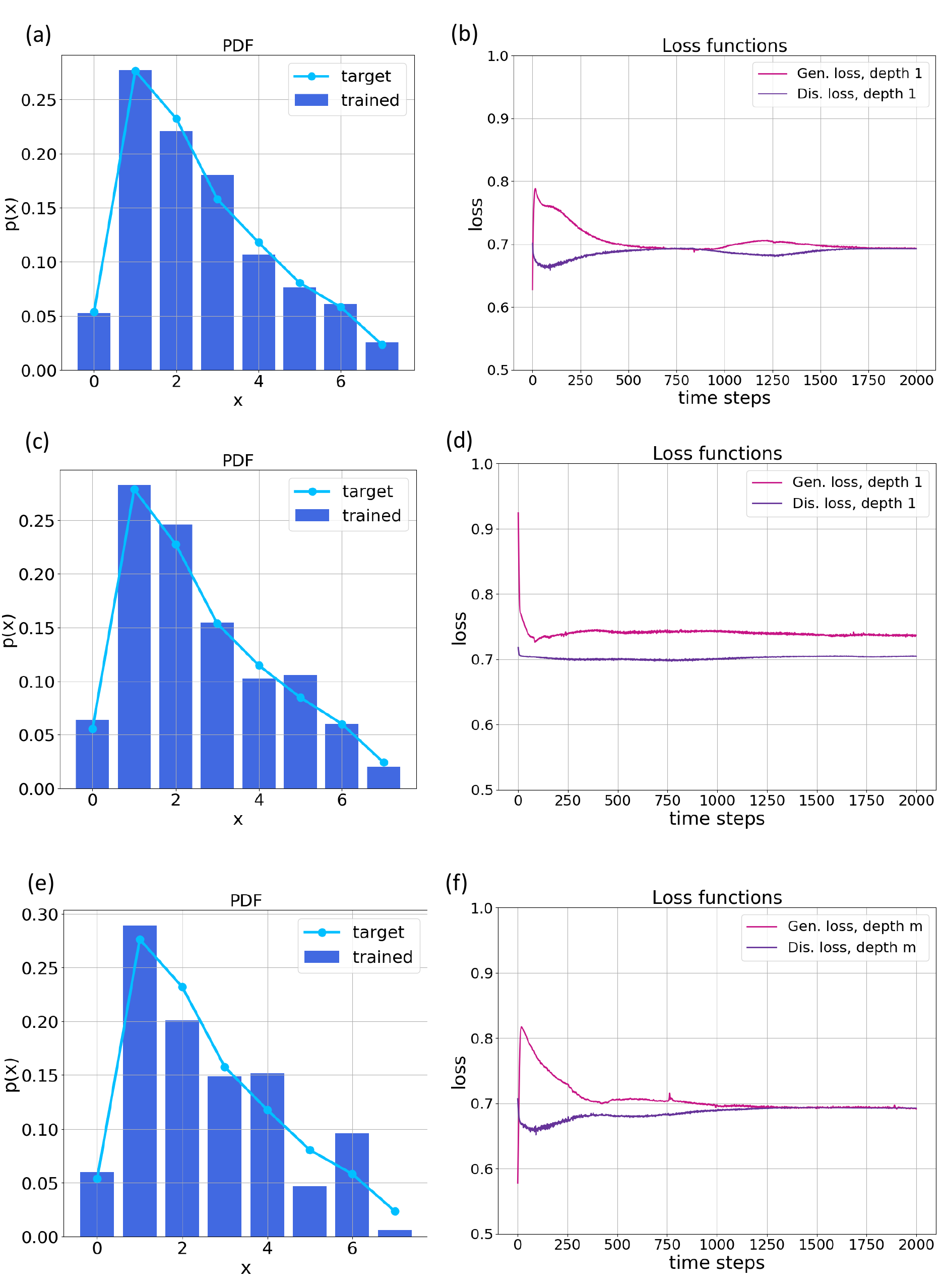}
	}

\caption{Results of training the qGAN with uniformly (a, b), normally (c, d), and randomly (e, f) initialized quantum generator. The PDFs corresponding to the trained $\ket{g_{\theta}}$ (a, c, e), as well as, the loss functions the generator and the discriminator (b, d, f) are illustrated.}
 \label{fig:training}
\end{figure}

First, we present results from running the qGAN training with a quantum simulator.
The training procedure involves $2,000$ epochs.
Fig.~\ref{fig:training} shows the progress of the loss functions $L_D\left(\phi, \theta\right)$ and $L_G\left(\phi, \theta\right)$, as well as, the probability density function (PDF) corresponding to the trained $\ket{g_{\theta}}$ and the target PDF.
The PDFs visualize that both uniform and normal initialization perform better than the random initialization.


\begin{figure}[h!]
\captionsetup{singlelinecheck = false, format= hang, justification=raggedright, font=footnotesize, labelsep=space} 
\centering{
\includegraphics[width=\linewidth]{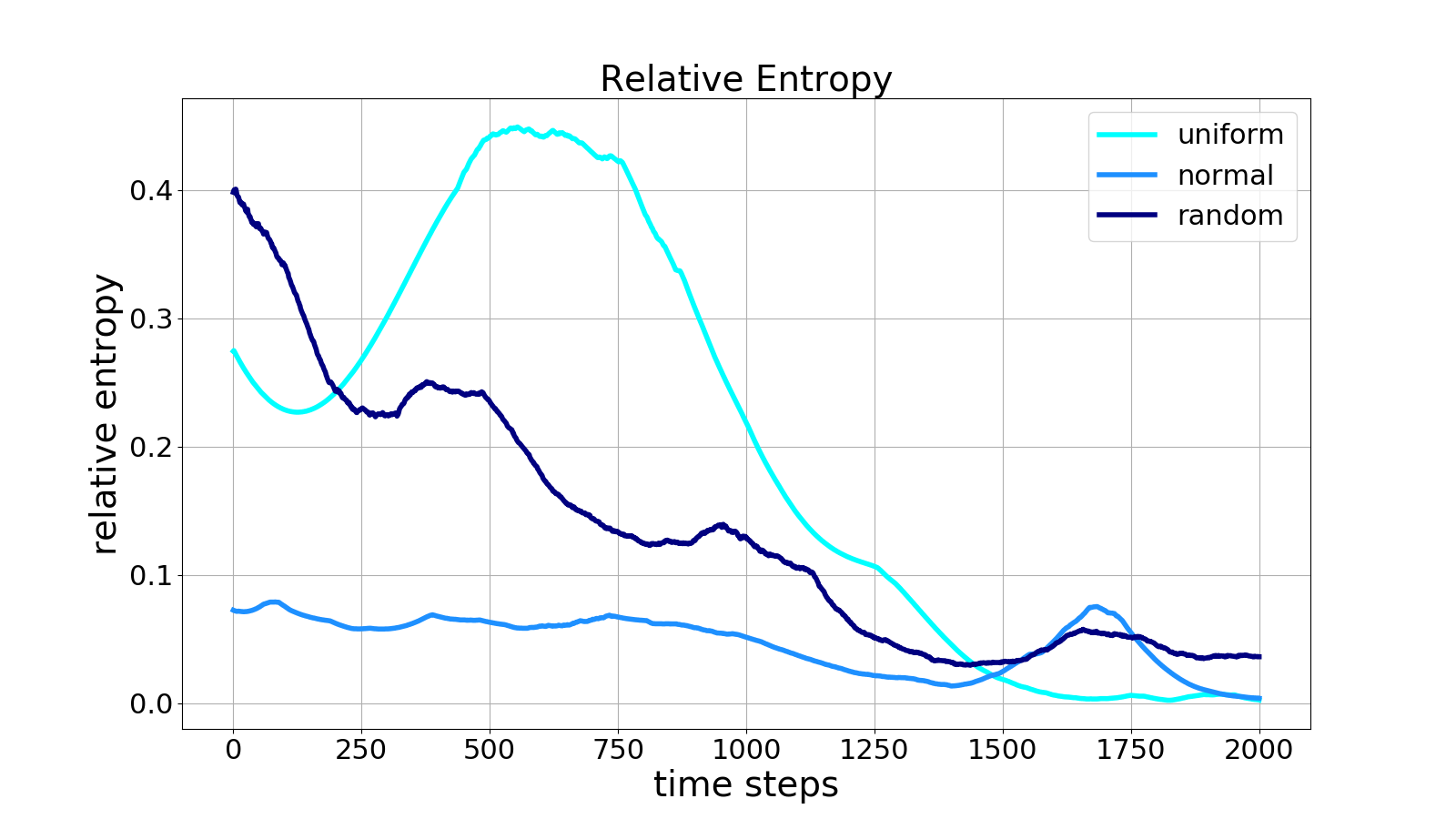}}
\caption{This figure illustrates the convergence of the relative entropy for $2,000$ training epochs with uniform, normal, and random initialization.}
\label{fig:rel_ent}
\end{figure}

Fig.~\ref{fig:rel_ent} shows the progress of the relative entropy and, thereby, illustrates how the generated distributions converge towards the training data's underlying distribution.
This figure also shows that the generator model which is initialized randomly performs worst.
Notably, the initial relative entropy for the normal distribution is already small.
We conclude that a carefully chosen initialization clearly improves the training, although all three approaches eventually lead to reasonable results.

Table~\ref{tbl:ks} presents the Kolmogorov-Smirnov statistics of the experiments.
The results also confirm that initialization impacts the training performance.
The statistics for the normal initialization are better than for the uniform initialization, which itself outperforms random initialization. It should be noted that the null-hypothesis is accepted for all settings.

\begin{table}[h!]
\captionsetup{singlelinecheck = false, format= hang, justification=raggedright, font=footnotesize, labelsep=space}
\begin{tabular}{ccc}
initialization &  $D_{KS}$ & Accept/Reject \\
 \hline
 uniform & $0.0369$ & Accept  \\
 normal  & $0.0320$ & Accept  \\
 random  & $0.0560$ & Accept
\end{tabular}
\caption{Kolmogorov-Smirnov statistic for randomly chosen samples from $\ket{g_{\theta}}$ and from the discretized, truncated log-normal distribution ${X}$.}
\label{tbl:ks}
\end{table}


Next, we present the results of the qGAN training run on an actual quantum processor, more precisely, the IBM Q Boeblingen chip \cite{ibmQX}. 
We use the same training data, quantum generator and discriminator as before. To improve the robustness of the training against the noise introduced by the quantum hardware, we set the optimizer's learning rate to $10^{-3}$. 
The initialization is chosen according to the random setting because it requires the least gates. 
Due to the increased learning rate, it is sufficient to run the training for $200$ optimization epochs.
For more details on efficient implementation of the generator on IBM Q Boeblingen, see Appendix \ref{app:HardwareEff}. 

Equivalent to the simulation, in each epoch, the training data is shuffled and split into batches of size $2,000$.
The generated data samples are created by preparing and measuring the quantum generator $2,000$ times.
To compute the analytic gradients for the update of $\theta$, we use $8,000$ measurement to achieve suitably accurate gradients.

\begin{figure}[h!]
\captionsetup{singlelinecheck = false, format= hang, justification=raggedright, font=footnotesize, labelsep=space} 
  	 \centering{
   \includegraphics[width=0.5\textwidth]{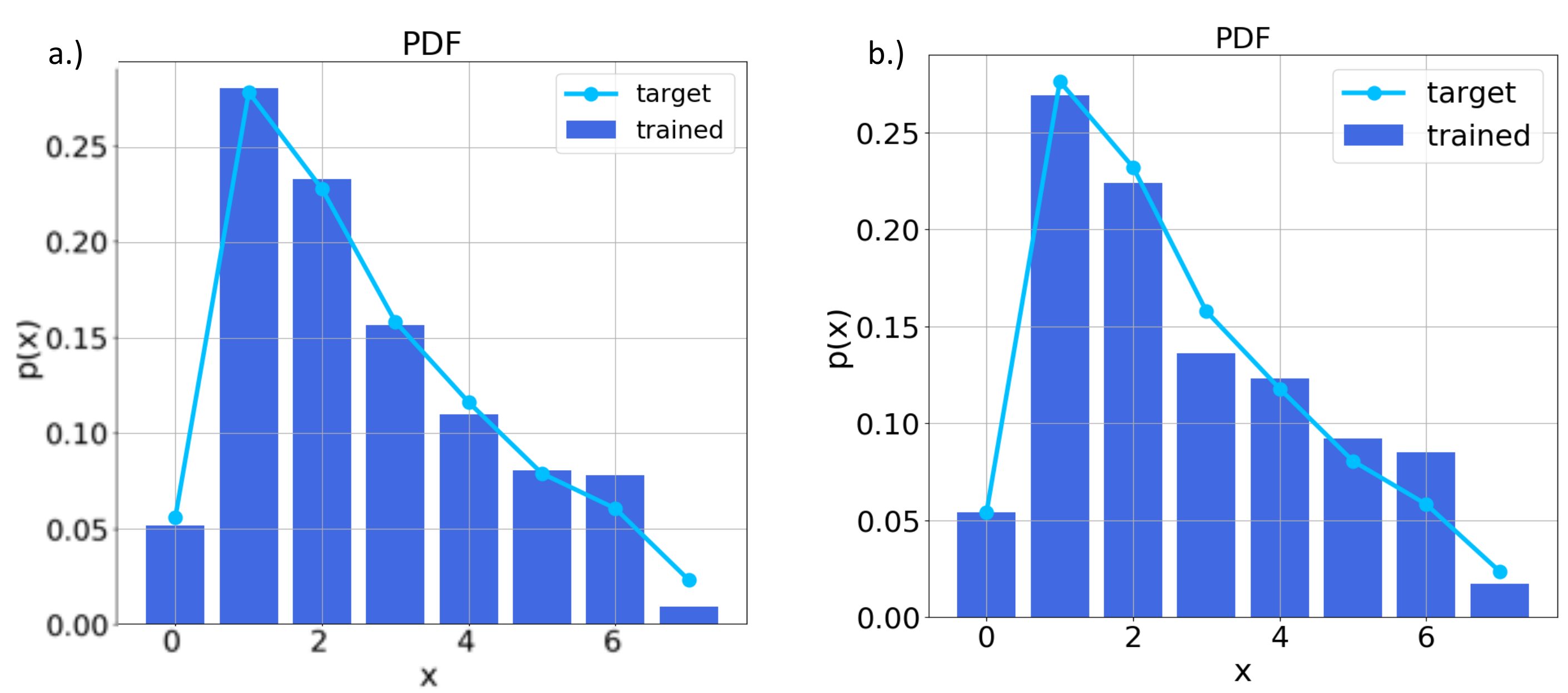}
   }
\caption{The presented PDFs from $\ket{g_{\theta}}$ are achieved with a randomly initialized qGAN training run on (a) the IBM Q Boeblingen and (b) a quantum simulation employing a noise model. }
 \label{fig:pdf_real}
\end{figure}

\begin{figure}[h!]
\captionsetup{singlelinecheck = false, format= hang, justification=raggedright, font=footnotesize, labelsep=space} 
  	 \centering{
   \includegraphics[width=0.5\textwidth]{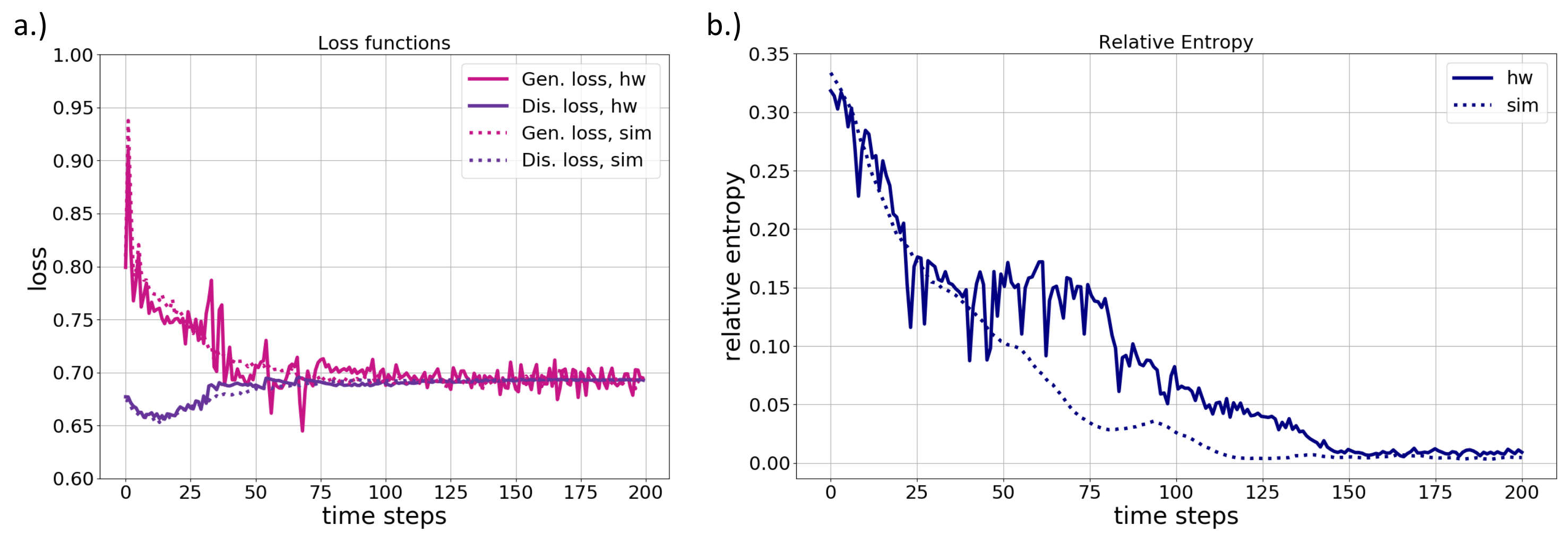}
   }
\caption{The figure illustrates the progress in the (a) loss functions and the (b) relative entropy during the training of the randomly initialized qGAN with the IBM Q Boeblingen quantum computer and a noisy quantum simulation. }
 \label{fig:rel_ent_real}
\end{figure}

Fig.~\ref{fig:pdf_real} presents the PDF corresponding to $\ket{g_{\theta}}$ trained with IBM Q Boeblingen respectively with a classical quantum simulation that models the quantum chip's noise.
To evaluate the training performance, we evaluate again the relative entropy and the Kolmogorov-Smirnov statistic.
A comparison of the progress of the loss functions and the relative entropy for a training run with the IBM Q Boeblingen chip and with the noisy quantum simulation is shown in Fig.~\ref{fig:rel_ent_real}.
The plot illustrates that the relative entropy for both, the simulation and the real quantum hardware, converge to values close to zero and, thus, that in both cases $\ket{g_{\theta}}$ evolves towards the random distribution underlying the training data samples.

Again, the Kolmogorov-Smirnov statistic $D_{KS}$ determines whether the null-hypothesis is accepted or rejected with a confidence level of $95\%$.
The results presented in Table~\ref{tbl:ks_hw} confirm that we were able to train an appropriate model on the actual quantum hardware.

\begin{table}[h!]
\captionsetup{singlelinecheck = false, format= hang, justification=raggedright, font=footnotesize, labelsep=space}
\begin{tabular}{cccc}
initialization & backend  &  $D_{KS}$ & Accept/Reject \\
 \hline
 random & simulation & $0.0420$ & Accept \\
 random & quantum computer & $0.0224$ & Accept

\end{tabular}
\caption{Kolmogorov-Smirnov statistic for randomly chosen samples of $\ket{g_{\theta}}$ trained with a noisy quantum simulation and using the IBM Q Boeblingen device.}
\label{tbl:ks_hw}
\end{table}

Notably, some of the more prominent fluctuations might be due to the fact that the IBM Q Boeblingen chip is recalibrated on a daily basis which is, due to the queuing, circuit preparation, and network communication overhead, shorter than the overall training time of the qGAN.

\subsection{European Option Pricing}
\label{subsec:results}

\begin{figure}[h!]
\captionsetup{singlelinecheck = false, format= hang, justification=raggedright, font=footnotesize, labelsep=space} 
\centering{
\includegraphics[width=\linewidth]{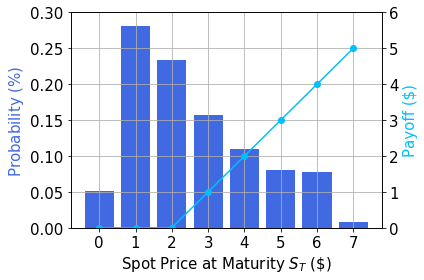}}
\caption{Probability distribution of the spot price at maturity $S_T$ and the corresponding payoff function for a European Call option.
The distribution has been learned with a randomly initialized qGAN run on the IBM Q Boeblingen chip.}
\label{fig:european_option}
\end{figure}

In the following, we demonstrate that the qGAN based data loading scheme enables the exploitation of potential quantum advantage of algorithms such as QAE by using a generator trained with actual quantum hardware to facilitate European call option pricing.
The resulting quantum generator loads a random distribution that approximates the spot price at maturity $S_T$.
More specifically, we integrate the distribution loading quantum channel into a quantum algorithm based on QAE to evaluate the expected payoff $\EX\left[\max\left\{S_T - K, 0 \right\}\right]$ for $K=\$2$, illustrated in Fig.~\ref{fig:european_option}.  
Given this efficient, approximate data loading, the algorithm can achieve a quadratic improvement in the error scaling compared to classical Monte Carlo simulation.
We refer to \cite{wor} and to Appendix~\ref{app:QAE} for a detailed discussion of derivative pricing with QAE.

The results for estimating $\EX\left[\max\left\{ S_T - K, 0 \right\}\right]$ are given in Tab.~\ref{tbl:europeanCallOption}, where we compare
\begin{itemize}
\item an analytic evaluation with the exact (truncated and discretized) log-normal distribution $p_{\text{real}}$,
\item a Monte Carlo simulation utilizing $\ket{g_{\theta}}$ trained and generated with the quantum hardware (IBM Q Boeblingen), i.e.~$1,024$ random samples of $S_T$ are drawn by measuring $\ket{g_{\theta}}$ and used to estimate the expected payoff, and

\item a classically simulated QAE-based evaluation using $m=8$ evaluation qubits, i.e.~$2^8=256$ quantum samples, where the probability distribution $\ket{g_{\theta}}$ was trained with IBM Q Boeblingen chip.

\end{itemize}

The resulting confidence intervals (CI) are shown for a confidence level of $95\%$ for the Monte Carlo simulation as well as the QAE.
The CIs are of comparable size, although, because of better scaling, QAE requires only a fourth of the samples.
Since the distribution is approximated, both CIs are close to the exact value but do not actually contain it.

Note that the estimates and the CIs of the MC and the QAE evaluation are not subject to the same level of noise effects. This is due to the fact, that the QAE evaluation uses the generator parameters trained with IBM Q Boeblingen but is run with a quantum simulator, whereas the Monte Carlo simulation is solely run on actual quantum hardware. To be able to run QAE on a quantum computer, further improvements are required, e.g.~longer coherence times and higher gate fidelities.

\begin{table}[h!]
\captionsetup{singlelinecheck = false, format= hang, justification=raggedright, font=footnotesize, labelsep=space}
\begin{tabular}{ccccc}
Approach & Distribution & Payoff (\$) & \#Samples  & CI (\$)\\ 
\hline
Analytic    & Log-normal       &  1.0602  & -    & - \\
MC + QC    & $\ket{g_{\theta}}$ & 0.9740  & 1024& $\pm 0.0848$ \\
QAE     & $\ket{g_{\theta}}$ &  1.1391  &  256 & $\pm 0.0710$
\end{tabular}
\caption{This table presents a comparison of different approaches to evaluate $\EX\left[\max\lbrace S_T - K, 0 \rbrace\right]$: an analytic evaluation of the log-normal model, a Monte Carlo simulation drawing samples from IBM Q Boeblingen, and a classically simulated QAE.
Furthermore, the estimates' $95\%$ confidence intervals are shown.}
\label{tbl:europeanCallOption}
\end{table}

\section{Conclusion and Outlook} \label{sec:discussion}
We demonstrated the application of an efficient, approximate probability distribution learning and loading scheme based on qGANs that requires $\mathcal{O}\left(poly\left(n\right)\right)$ many gates. In contrast to this, current state-of-the-art techniques for exact loading of generic random distributions into an $n$-qubit state necessitate $\mathcal{O}\left(2^n\right)$ gates which can easily predominate a quantum algorithm's complexity. 

The respective quantum channel is implemented by a gate-based quantum algorithm and can, therefore, be directly integrated into other gate-based quantum algorithms. 
This is explicitly shown by the learning and loading of a model for European call option pricing which is evaluated with a QAE-based algorithm that can achieve a quadratic improvement compared to classical Monte Carlo simulation.
Flexibility is given because the model can be fitted to the complexity of the underlying data and the loading scheme's resolution can be traded off against the complexity of the training data by varying the number of used qubits $n$ and the circuit depth $k$.
Moreover, qGANs are compatible with online or incremental learning, i.e.~the model can be updated if new training data samples become available.
This can lead to a significant reduction of the training time in real-world learning scenarios.


Some questions remain open and may be subject to future research, for example, an analysis of optimal quantum generator and discriminator structures as well as training strategies. 
Like in classical machine learning, it is neither apriori clear what model structure is the most suitable for a given problem nor what training strategy may achieve the best results.

Furthermore, although barren plateaus \cite{Clean_2018_BarrenPlateaus} were not observed in our experiments, the possible occurrence of this effect, as well as, counteracting methods should be investigated. Classical ML already offers a large variety of potential solutions, e.g.~ the inclusion of noise and momentum terms in the optimization procedure, the simplification of the function landscape by increase of the model size \cite{Stanford200_barrenplat} or the computation of higher order gradients. Moreover, schemes that were developed in the context of VQE algorithms, such as adaptive initialization \cite{Grimsley19AdaptiveVQE}, could help to circumvent this issue.

Another interesting topic worth investigating considers the representation capabilities of qGANs with other data types.
Encoding data into qubit basis states naturally induces a discrete and equidistantly distributed set of represented data values.
However, it might be interesting to look into the compatibility of qGANs with continuous or non-equidistantly distributed values.

\section{Acknowledgments}

We would like to thank Giovanni Mariani for sharing his knowledge and engaging in very helpful discussions.

IBM and IBM Q are trademarks of International Business Machines Corporation, registered in many jurisdictions worldwide.  Other product or service names may be trademarks or service marks of IBM or other companies.

\section{Code Availability}
The code for the qGAN algorithm is publicly available as part of Qiskit \cite{qiskit}. The algorithm can be found in \url{https://github.com/Qiskit/qiskit-aqua}. Tutorials explaining the training and the application in the context of QAE are located at \url{https://github.com/Qiskit/qiskit-iqx-tutorials}.

\appendix

\section{Isometric Quantum Generator} \label{app:isometry}

Closed quantum systems follow a unitary evolution.
The evolution of an open quantum system, i.e.~a quantum system that interacts with an environment, evolves according to an isometry instead of a unitary \cite{nielsen}.

In general, every isometry can be described by a unitary that acts on a larger system. In other words, an isometry is given by a partial trace of a unitary quantum state evolution.
The dynamics of an open quantum systems, and, thus, also a quantum generator acting as an isometry, can be implemented with additional ancilla qubits.
Depending on the setting, the use of an isometric quantum generator can be advantageous to learn random distributions, as mentioned in Sec.~\ref{sec:qGANs}.

\section{Multivariate Historical Data for Portfolio Optimization} \label{app:multi}

The qGAN scheme can also be used to learn and load multivariate random distributions.
Here, we present the learning and loading of a distribution underlying the first two principle components of multivariate, constant maturity treasury rates of US government bonds.
Note that the trained quantum channel can be used within the discussed QAE algorithm to evaluate, for instance, the fair price of a portfolio of government bonds, see \cite{wor}.

The following results are computed with a quantum simulation.
The training data set $X$ consists of more than $5,000$ samples, whereby data samples smaller than the $5\%-$percentile and bigger than the $95\%-$percentile have been discarded to reduce the number of required qubits for a reasonable representation of the distribution. 
The optimization scheme uses data batches of size $1,200$ and is run for $20,000$ training epochs.

Furthermore, we use depth $k \in \set{2, 3, 6}$, unitary quantum generators that act on $n=6$ qubits, i.e.~$3$ qubits per dimension (principle component).
The input state $\ket{\psi_{\text{in}}}$ is prepared as a multivariate uniform distribution and the generator parameters $\theta$ are initialized with random draws from a uniform distribution on the interval $[-\delta, +\delta]$ with $\delta = 10^{-1}$.

\begin{figure}[h!]
\captionsetup{singlelinecheck = false, format= hang, justification=raggedright, font=footnotesize, labelsep=space} 
\centering{
\includegraphics[width=\linewidth]{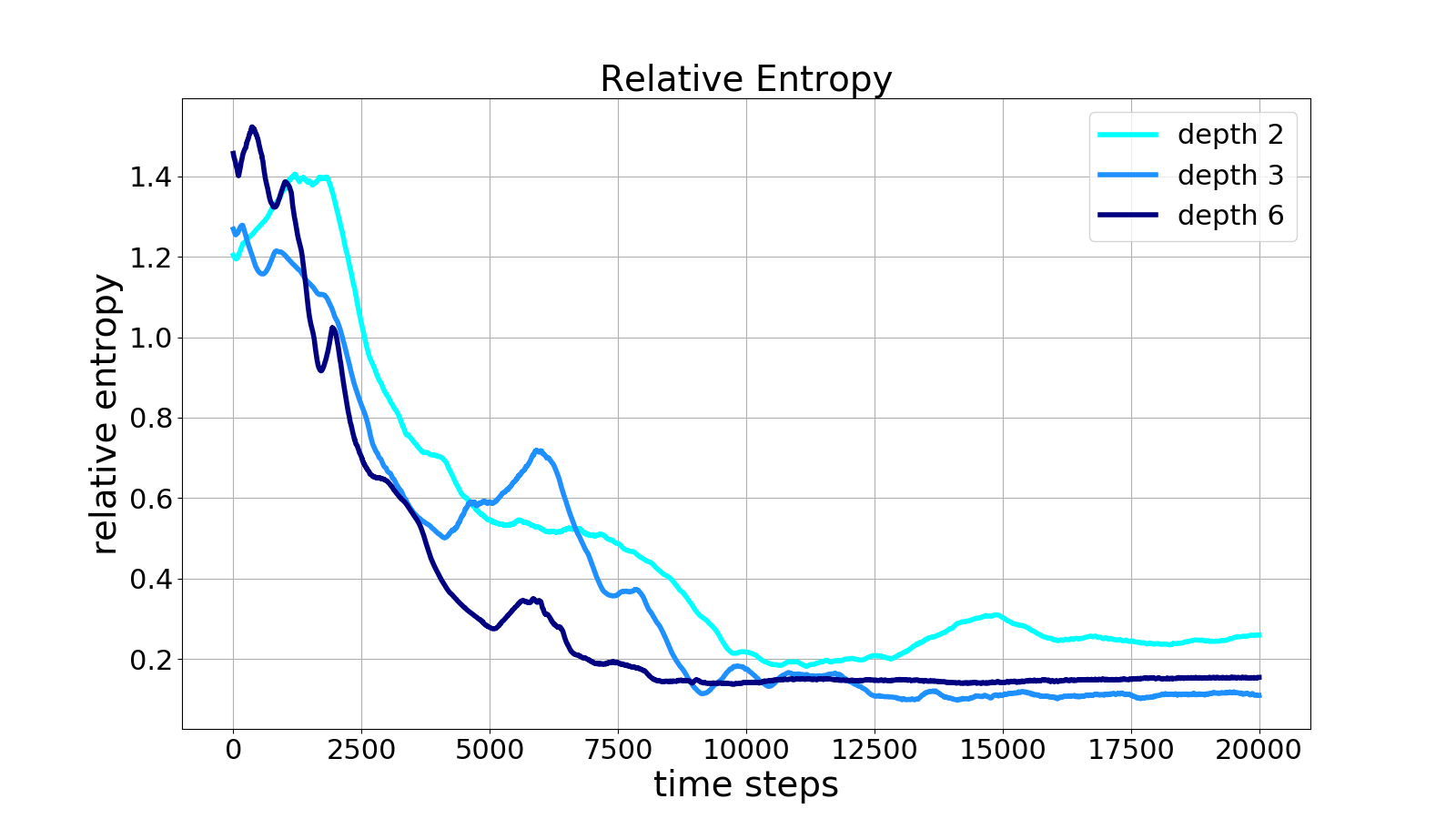}
}
\caption{The progress of the relative entropy between the quantum generator and the multivariate random distribution underlying the training data for depth $k \in \set{2, 3, 6}$.}
 \label{fig:rel_ent_multi}
\end{figure}

Here, the classical discriminator is composed of a $512-$ node input layer, a $256-$ node hidden-layer, and a single-node output layer. 
Equivalently to the discriminator described in Sec.~\ref{subsec:benchmarking}, the hidden layers apply linear transformations followed by Leaky ReLU functions \cite{Pedamonti_2018_Non-linear_Activation_Functions_NN} and the output layer employs a linear transformation followed by a sigmoid function.
The evolution of the relative entropy between the generated and the real probability distribution is shown Fig.~\ref{fig:rel_ent_multi}.

\section{Practical Initialization of a Normal Distribution} \label{app:normalInit}

\begin{figure}[h!]
\captionsetup{singlelinecheck = false, format= hang, justification=raggedright, font=footnotesize, labelsep=space}
\begin{center}
\includegraphics[width=0.5\textwidth]{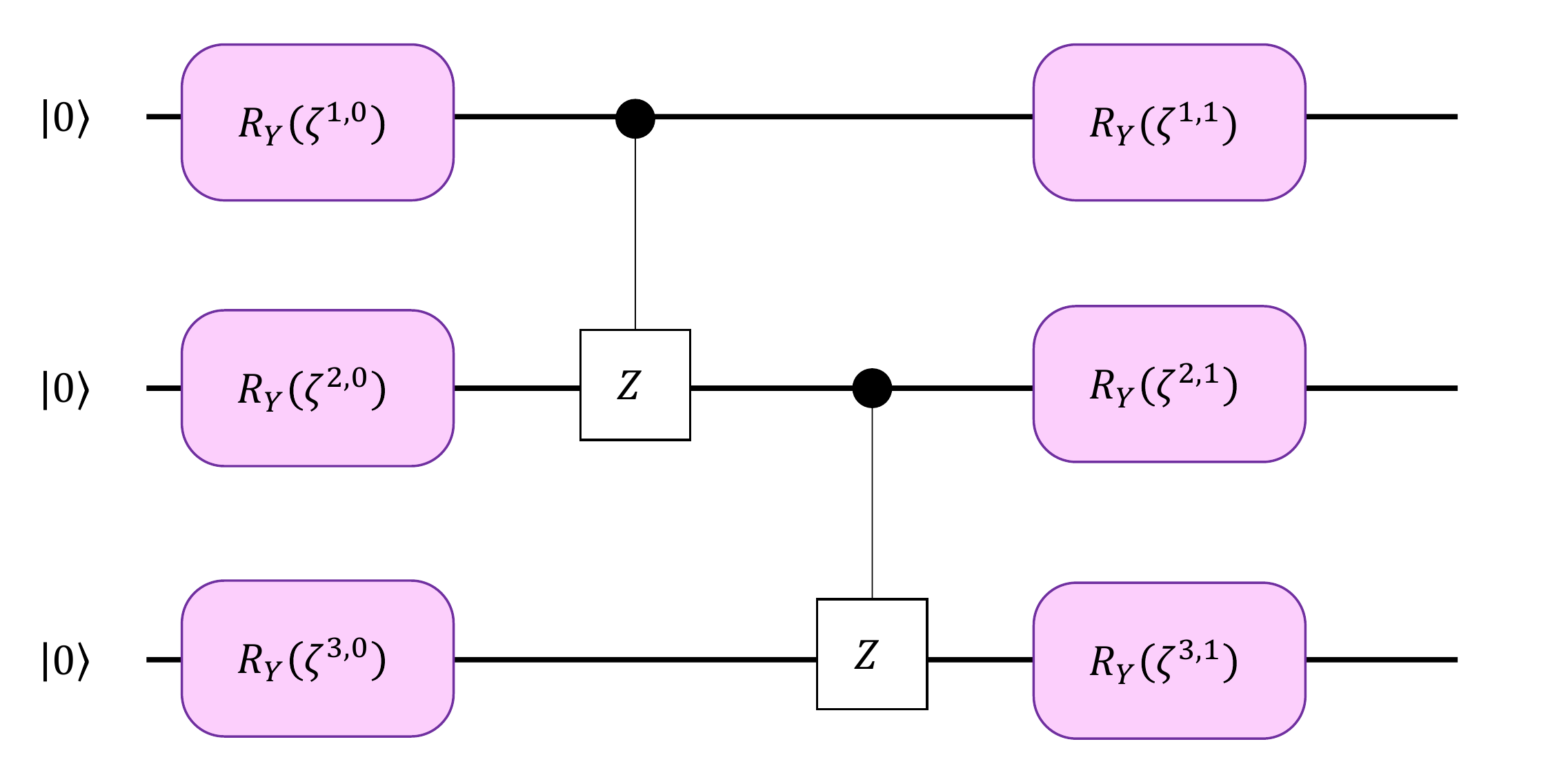}
\end{center}
\caption{Variational quantum circuit for approximate loading of a discretized normal distribution.}
\label{fig:normal_init}
\end{figure}

As proven in \cite{grover}, a normal distribution can be efficiently loaded into a quantum state.
However, the suggested loading method requires the use of involved quantum arithmetic techniques. 
Considering the illustrative examples from Sec.~\ref{subsec:benchmarking} and Sec.~\ref{subsec:lognormal}, it is sufficient to load an approximate normal distribution as initialization state.
This can be achieved by fitting the parameters of a $3$-qubit variational quantum circuit with depth $1$ with a least squares loss function.
More specifically, we minimize the distance between the measurement probabilities $p_{\zeta}^{i}$ of the circuit output and the probability density function of a discretized normal distribution $q^i$
\begin{equation}
	\min_{\zeta} \sum\limits_i \left\| p_{\zeta}^{i} - q^i \right\|^2.
\end{equation}
The circuit used for training is depicted in Fig.~\ref{fig:normal_init}.
Note that this approach does not scale, particularly not for higher-dimensional distributions.
The sole purpose of this approach is to generate shallow testing circuits.

The circuit's parameters were trained to approximate normal distributions with the mean and the standard deviation of the data samples drawn from each a log-normal distribution with $\mu = 1$ and $\sigma = 1$,
\begin{equation*}
	\theta_\text{{lognormal}}= \left[0.3580, 1.0903, 1.5255, 1.3651, 1.4932, -0.9092\right], 
\end{equation*}
a triangular distribution with lower limit $l=0$, upper limit $u=7$ and $\mu=2$,
\begin{equation*}
\begin{split}
	\theta_\text{{triangular}} = \left[1.5343, 1.6183, 0.8559, -0.4041, 0.4953, 1.2238\right]
	\end{split}
\end{equation*}
 and a bimodal distribution consisting of two superimposed Gaussian distributions with $\mu_1=0.5$, $\sigma_1=1$ respectively $\mu_2=3.5$, $\sigma_2=0.5$,
 \begin{equation*}
 	\theta_\text{{bimodal}} = \left[0.4683, 0.8200, 1.4512, 1.1875, 1.3883, -0.8418\right]
 \end{equation*}, whereby the least square errors are of the order $10^{-4}$.

\section{Analytic Gradients of a Variational Quantum Circuit} \label{app:gradients}

Compared to gradient-free optimization, gradient-based optimization methods have the potential to improve convergence rates, e.g. in a convex vicinity of local optima \cite{Harrow2019_QGradients}. We now discuss a method to calculate analytic gradients \cite{Farhi2018_gradients, Schuld2018CircuitcentricQC, Fujii2018_qcircuitLearn,  LearnandInference, DifferentiableLearning} for the variational circuit illustrated in Fig.~\ref{fig:varForm}.

Applying our $n-$qubit generator to the input state gives
\begin{equation}
\label{eq:state}
\begin{split}
\ket{g_{\theta}} &= G_{\theta}\ket{\psi_{\text{in}}} \\
&= \prod\limits_{p=1}^{k}\left(\bigotimes\limits_{q=1}^{n}\left(R_Y\left(\theta^{q,p}\right) \right)U_{\text{ent}} \right) \bigotimes\limits_{q=1}^{n}\left(R_Y\left(\theta^{q,0}\right)\right)\ket{\psi_{\text{in}}} \\
&=  \sum\limits_{j=0}^{2^n-1}\sqrt{p_{\theta}^{j}}\ket{j}.
	\end{split}
\end{equation}

We measure $\ket{g_{\theta}}\quad m$ times to obtain data samples $g^l, \: l\in\set{1, \ldots, m}$ which can take $2^n$ different values. The generator loss function for a data batch of size $m$ reads
\begin{align}
\begin{split}
	 L_G\left(\phi, \theta\right) = -\frac{1}{m}\sum\limits_{l=1}^{m}\log\left(D_{\phi}\left(g^{l}\right)\right),
	\end{split}
\end{align}
or equivalently,
\begin{align}
\begin{split}
	 L_G\left(\phi, \theta\right) =-\sum\limits_{j=0}^{2^n-1} p_{\theta}^j\log\left(D_{\phi}\left(g^{j}\right)\right),
	\end{split}
\end{align}
with 
\begin{align}
	p_{\theta}^j = \vert\braket{j\vert g_{\theta}}\vert^2.
\end{align}

Updating the parameters $\theta$ with gradient based methods requires the evaluation of

\begin{equation}
\label{eq:derivativeLossG}
\begin{split}
\frac{\partial L_G\left(\phi,\theta\right)}{\partial \theta^{i,l}} &= - \sum\limits_{j=1}^{m}\frac{\partial p_{\theta}^j}{\partial \theta^{i,l}} \log\left(D_{\phi}\left(g^j\right)\right).
\end{split}
\end{equation}

According to \cite{Fujii2018_qcircuitLearn} Eq.~\eqref{eq:derivativeLossG} can be evaluated by
\begin{equation}
\label{eq:prob_derivative_pi}
\begin{split}
\frac{\partial p_{\theta}^j}{\partial \theta^{i,l}} = \frac{1}{2}\left(p_{\theta_{+}^{i,l}}^j - p_{\theta_{-}^{i,l}}^j \right),
	\end{split}
\end{equation}
with $\theta_{\pm}^{i,l} = \theta^{i,l} \pm \frac{\pi}{2}e_{i,l}$ and $e_{i,l}$ denoting the $(i,l)$-unit vector of the respective parameter space.

\section{Statistical Measures} \label{app:statMeas}

Two different statistical measures are utilized to evaluate the performance of the qGAN.
Both measures are defined as a distance of two (empirical) probability distributions $P$ and $Q$.

The Kolmogorov-Smirnov statistic \cite{kolmogorov, JUSTEL1997_multiKS} is based on the (empirical) cumulative distribution functions $P\left(X\leq x\right)$ and $Q\left(X\leq x\right)$ and is given by
\begin{equation}
\label{eq:KS}
D_{KS}\left(P||Q\right)  = \underaccent{x\in X}{sup}\:\vert P\left(X\leq x\right) - Q\left(X\leq x\right)\vert.
\end{equation}

The statistic can be used as a goodness-of-fit test.
Given the null-hypothesis $P\left(x\right) = Q\left(x\right)$, we draw $s=500$ samples from both distributions and choose a confidence level $(1 - \alpha)$ with $\alpha = 0.05$.
The null-hypothesis is accepted if 
\begin{equation}
D_{KS}\left(P\Vert Q\right)  \leq \sqrt{\frac{\ln{\frac{2}{\alpha}}}{s} } = 0.0859. 
\end{equation}

Another measure that can be used to characterize the closeness of (empirical) discrete probability distributions $P\left(x\right)$ and $Q\left(x\right)$ is the relative entropy, also called Kullback-Leibler divergence \cite{nielsen,kullback1951}.
This entropy-related measure is given by
\begin{equation}
\label{eq:relEntr}
D_{RE}\left(P||Q\right) = \sum\limits_{x\in X}P\left(x\right)\log\left(\frac{P\left(x\right)}{Q\left(x\right)}\right).
\end{equation}
The relative entropy represents a non-negative quantity, i.e.~$D_{RE}\left(P||Q\right)\geq 0$, where $D_{RE}\left(P||Q\right) = 0$ holds if $P\left(x\right) = Q\left(x\right),$ for all values $x$.

\section{Quantum Amplitude Estimation}
\label{app:QAE}
\begin{figure}[h!]
\captionsetup{singlelinecheck = false, format= hang, justification=raggedright, font=footnotesize, labelsep=space}
\begin{center}
\includegraphics[width=\linewidth]{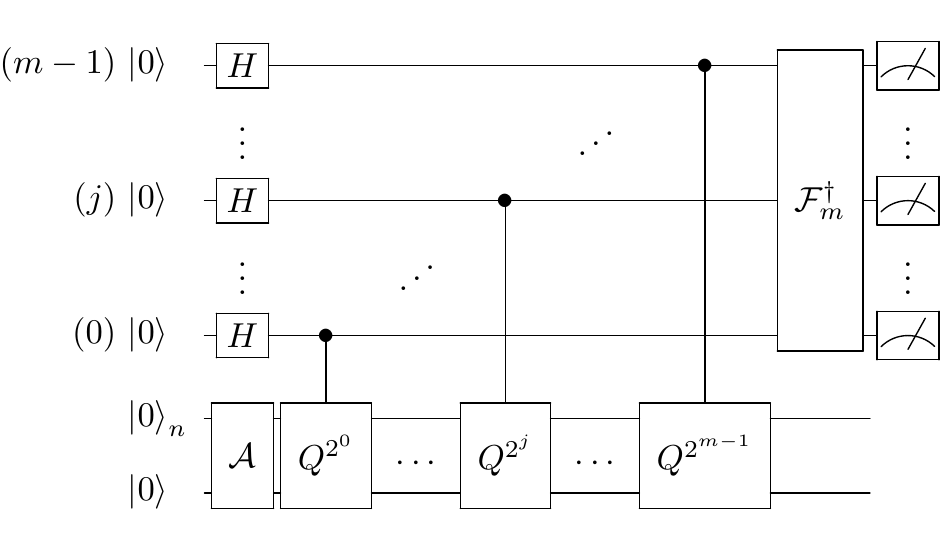}
\end{center}
\caption{The quantum circuit corresponding to Quantum Amplitude Estimation algorithm with the inverse Quantum Fourier Transform \cite{nielsen} being denoted by $\mathcal{F}_m^{\dagger}$.}
\label{fig:qae}
\end{figure}

Given a quantum channel 
\begin{equation}
\label{eq:qae}
\mathcal{A}\ket{0}^{\otimes n+1} = \sqrt{1-a}\ket{\psi_0}\ket{0} + \sqrt{a}\ket{\psi_1}\ket{1},
\end{equation}
where $\ket{\psi_0}$, $\ket{\psi_1}$ denote $n$-qubit states, the QAE algorithm \cite{brassard}, illustrated in Fig.~\ref{fig:qae}, enables the efficient evaluation of the amplitude $a$.
The algorithm requires $m$ additional evaluation qubits that control the applications of an operator $\mathcal{Q} = - \mathcal{A}\mathcal{S}_0\mathcal{A}^{\dagger}\mathcal{S}_{\psi_0}$ where $\mathcal{S}_0 = \mathbb{I}^{\otimes n +1} - 2\ket{0}\bra{0}^{\otimes n+1}$ and $\mathcal{S}_{\psi_0} = \mathbb{I}^{\otimes n +1} - 2\ket{\psi_0}\bra{\psi_0}\otimes\ket{0}\bra{0}$.

The error in the outcome - ignoring higher terms - can be bounded by $\frac{\pi}{2^m}$. Considering that $2^m$ is the number of quantum samples used for the estimate evaluation, this error scaling is quadratically better than the classical Monte Carlo simulation.

To use QAE for the pricing of European options, we need to construct and implement a suitable oracle $\mathcal{A}$. First, we load the uncertainty distribution that represents the spot price $S_T$ of the underlying asset at the option's maturity $T$ into a quantum state $\sum_{i=0}^{2^n-1} \sqrt{p_i}\ket{i}$. It should be noted that small errors in this state preparation only lead to small errors in the final result.
Then, we add an ancilla qubit $\ket{0}$ and use a comparator circuit which applies an $X$ gate to the ancilla if $i>K$, i.e.
\begin{eqnarray}
\ket{i}\ket{0} \mapsto \begin{cases}
\ket{i}\ket{0} & \text{, if} \;  i \leq K \\
\ket{i}\ket{1} &  \text{, if} \; i > K,
\end{cases}
\end{eqnarray}
where $K$ denotes the strike price. 
Now, the state reads
\begin{equation}
\sum_{i=0}^{K} \sqrt{p_i} \ket{i}\ket{0} + \sum_{i=K+1}^{2^n-1} \sqrt{p_i} \ket{i} \ket{1}
\end{equation}
Finally, we control the mapping of the payoff function to the amplitude of another ancilla qubit $\ket{0}$ with the comparison ancilla.
This construction implements channel $\mathcal{A}$ and approximates the quantum state
\begin{eqnarray}
\label{eq:qaeRot}
\begin{split}
\mathcal{A}\ket{0}^{\otimes n+1} = \sum_{i=0}^{K} \sqrt{p_i} \ket{i}\ket{0} \ket{0}+ \\
\sum_{i=K+1}^{2^n-1} \sqrt{p_i} \ket{i} \ket{1} \left( \sqrt{1 - f(i)} \ket{0} + \sqrt{f(i)} \ket{1} \right),
\end{split}
\end{eqnarray}
with $f(i) = \frac{i - K}{2^n - K - 1}$.
For practical reasons, we avoid the involved implementation of the exact linear objective rotation given in Eq.~\eqref{eq:qaeRot} by applying the approximation scheme introduced in \cite{wor}.

Eventually, the probability of measuring $\ket{1}$ in the last ancilla is equal to
\begin{eqnarray}
\mathbb{P}[\ket{1}] &=&
\frac{1}{2^n - K - 1} \sum_{i=K+1}^{2^n-1} p_i (i - K) \\
&=& \frac{1}{2^n - K - 1} \mathbb{E}[\max\{0, S_T - K\}].
\end{eqnarray}
We can see from comparing Eq.~\eqref{eq:qae} and Eq.~\eqref{eq:qaeRot} that $\mathbb{P}[\ket{1}] = a$.
It follows that we can use QAE to efficiently evaluate $\mathbb{E}[\max\{0, S_T - K\}] = \mathbb{P}[\ket{1} ](2^n - K - 1)$.

\section{Hardware Efficient Circuit Implementation}
 \label{app:HardwareEff}

\begin{figure}[h!]
\captionsetup{singlelinecheck = false, format= hang, justification=raggedright, font=footnotesize, labelsep=space} 
  	 \centering{
   \includegraphics[width=0.3\textwidth]{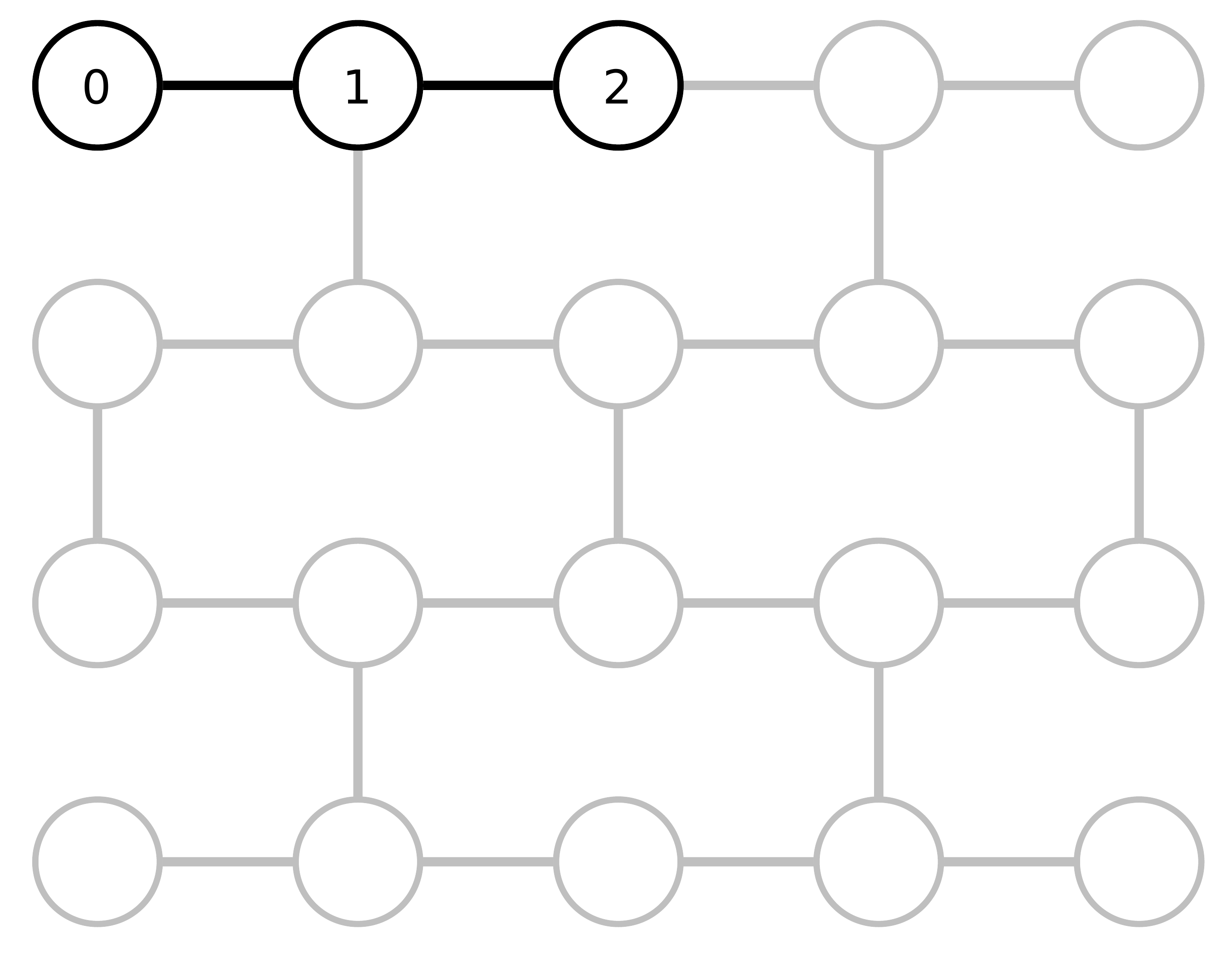}
   }
\caption{The figure illustrates the connectivity of the IBM Q Boeblingen 20 superconducting qubit chip, as well as, the qubits  used for the qGAN training.}
 \label{fig:bobby}
\end{figure}

\begin{figure}[h!]
\captionsetup{singlelinecheck = false, format= hang, justification=raggedright, font=footnotesize, labelsep=space} 
  	 \centering{
   \includegraphics[width=0.4\textwidth]{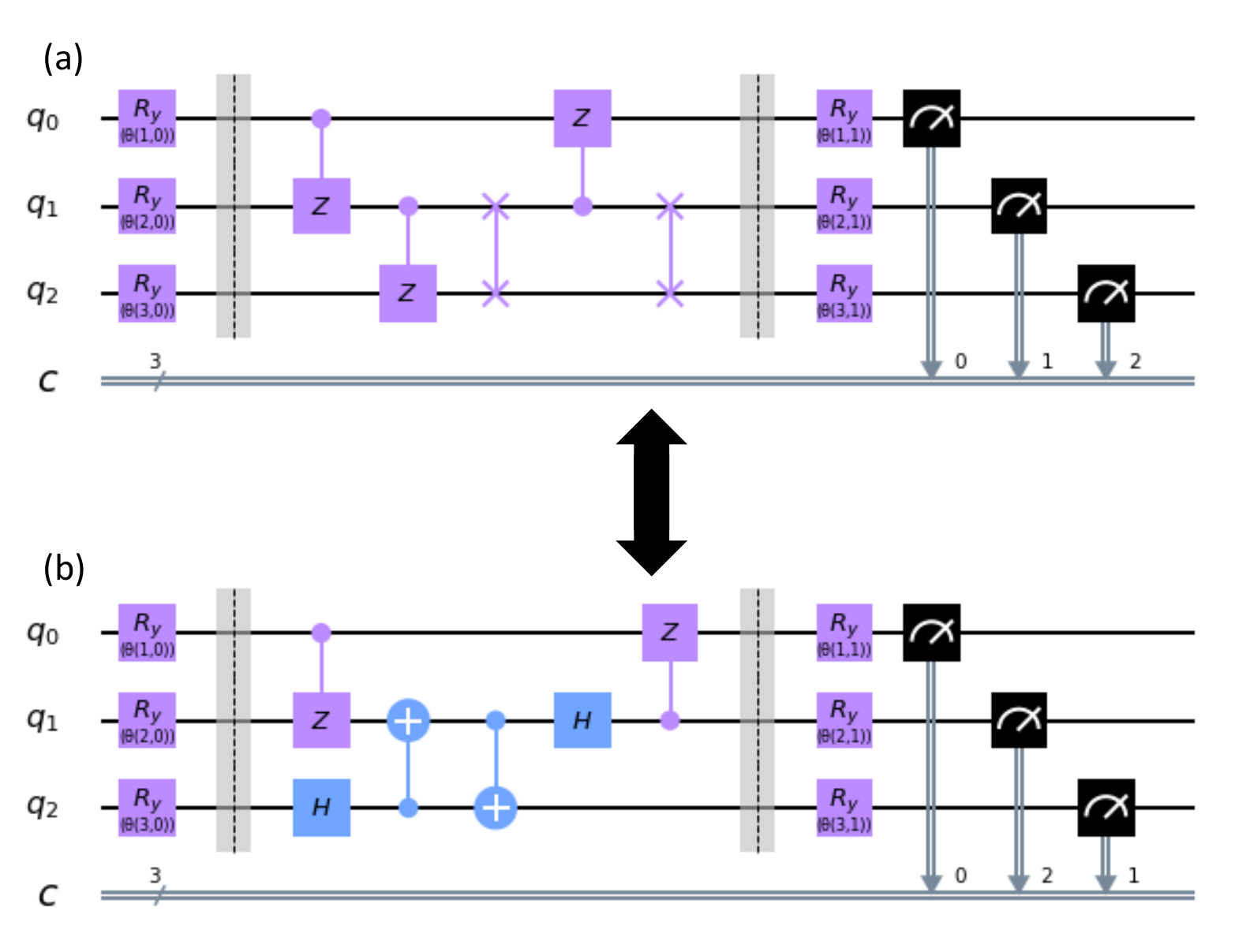}
   }
\caption{The action of the illustrated circuits is equivalent. Since the quantum circuit at the bottom requires fewer $CX$ gates, it is the favorable implementation choice for training a qGAN with actual quantum hardware. Notably, the lower circuit projects the measurement of qubit $\text{q}_1$ ($\text{q}_2$) on bit $\text{c}_2$ ($\text{c}_1$).}
 \label{fig:hw_eff}
\end{figure}

Due to the connectivity layout of the IBM Q Boeblingen chip, shown in Fig.~\ref{fig:bobby}, any subset of three qubits - we use qubits 0, 1, 2 - has linear connectivity only. Thus, the implementation of the entanglement block presented in Fig.~\ref{fig:varForm} requires the use of SWAP gates, as shown in Fig.~\ref{fig:hw_eff}(a).
The implementation of $CZ\circ SWAP$ with the gate set currently available for IBM Q backends requires the use of $4\; CX$ gates, i.e.~$3$ for the $SWAP$ and $1$ for the $CZ$.
However, we can reduce the number of required $CX$ gates, see Fig.~\ref{fig:hw_eff}(b). As shown in  \cite{Vatan2004}, the action of circuit (a) is equivalent to the action of circuit (b), which only utilizes $2\; CX$ gates. During the training, circuit (b) maps the measurement of $\text{q}_1$ ($\text{q}_2$) on bit $\text{c}_2$ ($\text{c}_1$) to compensate for the second $SWAP$ in circuit (a). However, when using the generator circuit for data loading in another algorithm, such as QAE, an actual $SWAP$ gate must be implemented.

\pagebreak
\bibliographystyle{IEEEtranN}
\bibliography{references}

\end{document}